\documentclass[useAMS,usenatbib]{mn2e}

\def\simgt{\mathrel{\spose{\lower 3pt\hbox{$\sim$}}
        \raise 2.0pt\hbox{$>$}}}
\def\simlt{\mathrel{\spose{\lower 3pt\hbox{$\sim$}}\raise 2.0pt\hbox{$<$}}}

\title[Properties of High Redshift Quasars]{Properties of High Redshift Quasars-II: What does the quasar luminosity function tell us about super-massive black-hole evolution?}

\author[Wyithe \& Padmanabhan]{J. Stuart B. Wyithe$^{1}$, T. Padmanabhan$^{2}$\\
$^1$ School of Physics, University of Melbourne, Parkville, Victoria, Australia\\
$^2$ Inter-University Center for Astronomy and Astrophysics, Pune, India\\
 Email: swyithe@physics.unimelb.edu.au, nabhan@iucaa.ernet.in}

\date{Accepted Received}
\pagerange{\pageref{firstpage}--\pageref{lastpage}}
\pubyear{2005}

\def\LaTeX{L\kern-.36em\raise.3ex\hbox{a}\kern-.15em
    T\kern-.1667em\lower.7ex\hbox{E}\kern-.125emX}

\begin{document}

\label{firstpage}

\maketitle

\begin{abstract}
\noindent 
In the local universe, the masses of Super-Massive Black-Holes (SMBH)
appear to correlate with the physical properties of their hosts,
including the mass of the dark-matter halo. Using these clues as a
starting point many studies have produced models that can explain
phenomena like the quasar luminosity function. The shortcoming of this
approach is that working models are not unique, and as a result it is
not always clear what input physics is being constrained. Here we take
a different approach. We identify critical parameters that describe
the evolution of SMBHs at high redshift, and constrain their parameter
space based on observations of high redshift quasars from the Sloan
Digital Sky Survey. We find that the luminosity function taken in
isolation is somewhat limited in its ability to constrain SMBH
evolution due to some strong degeneracies. This explains the presence
in the literature of a range of equally successful models based on
different physical hypotheses. Including the constraint of the local
SMBH to halo mass ratio breaks some of the degeneracies, and our
results suggest halo masses at $z\sim4.8$ of $10^{12.5\pm0.3}M_\odot$
(with 90\% confidence), with a SMBH to halo mass ratio that decreases
with time ($>99\%$). We also find a quasar luminosity to halo mass
ratio that increases with halo mass ($>99\%$). These features need to be
incorporated in all successful models of SMBH evolution. On the other hand
current observations do not permit any conclusions regarding the
evolution of quasar lifetime, or the SMBH occupation fraction in dark
matter halos.
\end{abstract}


\section{Introduction}

The Sloan Digital Sky Survey (SDSS) has discovered luminous quasars at
redshifts as high as $z\sim6.4$, i.e., when the universe was only a
billion years old. The super-massive black-holes (SMBH) powering these
quasars have been estimated to have masses of $\sim10^9$ solar
masses. There is evidence that SMBHs contained a larger fraction of
the host bulge mass at high redshift (Merloni, Rudnick \& Di
Matteo~2004; Shields, Salviander, Bonnin~2006; Peng et al.~2006). There are, however,
several other questions regarding the galaxies that host these high
redshift quasars, and the physics that govern their evolution that
have remained largely unanswered. Traditionally, investigation of the
evolution of high redshift quasars has revolved around construction of
models for the luminosity function which can be compared with
observation. In this way successful models can be labeled as possible
candidates to explain the formation and evolution of the quasars. An
equally valuable, though a rarer approach is to rule out potential
models because they are unable provide an explanation for the observed
population. This approach is important because while a successful
model can only suggest a mechanism as one possibility, a model which
does not reproduce an observed phenomena may be definitely ruled out.

Many successful models have been produced to explain the high redshift
luminosity function (e.g. Haiman \& Loeb~1998; Haehnelt, Natarajan \&
Rees~1998; Kauffmann \& Haehnelt~2000; Volonteri, Haardt \&
Madau~2003; Wyithe \& Loeb~2003; Di~Matteo, Croft, Springel \&
Hernquist~2003). Some authors advocate the view that the simpler a
model the better because simple models can more clearly elucidate the
important underlying physics. Others would argue that more detailed
numerical models that attempt to capture more of the complex
non-linear processes involved offer a more realistic and reliable
description.  Models for the luminosity function, therefore, take a
variety of forms and complexity. These range from analytic models in
which simple physical prescriptions are combined with the
Press-Schechter~(1974) mass function and merger rates; to more complex
semi-analytic models with merger trees and parameterised laws to
describe physical processes; through to numerical simulations with
quasar activity included through a semi-analytic prescription. The
published models of all types are broadly successful in their
reproduction of high redshift number counts. In general this is
because if one adds the number of free parameters to the number of
physical assumptions, then one gets a number of parameters that is
comparable to or smaller than the effective number of constraints
(which in practice includes only the slope of the luminosity function
and the evolution of slope with redshift, in addition to the density
normalisation).
 
In this paper we highlight the fact that many different models are
able to reproduce the data on the high redshift quasar luminosity
function, and note that this implies some degeneracy among input
physical parameters or mechanisms. Indeed, one such degeneracy was
pointed out in an early model for high redshift quasar evolution
(Haehnelt, Natarajan \& Rees~1998). Assuming that quasar number counts
trace the number of dark-matter halo hosts, Haehnelt et al.~(1998)
found that the luminosity function at $z>2$ could be equally well
described using a constant SMBH to halo mass ratio combined with a
constant lifetime; or using a mass dependent SMBH to halo mass ratio
combined with a lifetime that decreased towards high redshift. The
latter possibility is expected from a feedback scenario for SMBH
growth (Silk \& Rees~1998). The degeneracy described above would imply
that one cannot obtain strong evidence for feedback using only the
high redshift luminosity function data. More generally, these types of
degeneracies mean that it is difficult to know which physical features
of the various models are being constrained by the data. Indeed we
would like to know whether it is possible to conclude anything from
published models other than the fact that the population of high
redshift quasars can be reconciled with the $\Lambda$CDM cosmology.

One feature that all models have in common is that they are based on
the density and evolution of the dark-matter halo population. In
semi-analytic models this population is generally modeled according to
the Press-Schechter~(1974) mass function (with extensions). Similar
results are obtained in cases where cosmological N-body codes are used
because the Press-Schechter mass function provides an accurate
reflection of simulations over the mass-range of interest. In common
with previous work, we use the Press-Schechter mass function as a
starting point from which the quasar population is generated. However
we do not attempt to model physical processes such as SMBH growth and
feedback etc., as is done in traditional semi-analytic
modeling. Instead we take the more general approach of applying
arbitrary parameterisations of processes like the evolution of quasar
lifetime, and the evolution of the relation between SMBH mass, halo
mass and redshift.  While it is true that we assign free parameters to
describe processes that we do not understand, the procedure for
finding {\em sets} of these parameters which satisfy the data in a
statistical sense is fundamentally different from the usual procedure
of semi-analytic modeling which aims to find a single successful
model. We are interested in the full range of allowed parameters as
well as which sets of parameters can be excluded, rather than in a
single set of parameters that is able to describe the data. In this
way we hope to provide guidance for future attempts to build physical
models of SMBH evolution.

The optical quasar luminosity function shows a peak in its evolution
at $z\sim2-3$. At higher redshifts the quasar population grows with
time, and it is natural to relate the rise of the quasar population to
the rise of the dark-matter halo population (e.g. Haehnelt, Natarajan
\& Rees 1998; Haiman \& Loeb~1998; Volonteri, Haardt \& Madau~2003;
Wyithe \& Loeb~2003).  However near a redshift of $z\sim2-3$ the
non-linear mass scale moves from galactic scale to cluster scale.  It
is thought that the rapid fall in the density of bright quasars below
$z\sim2$ is due to a combination of a dwindling supply of cold gas at
late times (Kauffmann \& Haehnelt~2000) with an injection of feedback
from the quasars into the surrounding IGM that prevents further gas
accretion onto collapsing systems (Scannapiecco
\& Oh~2004). Following the peak of quasar evolution one can no-longer
relate the growth of the quasar (or galaxy) population directly to
that of the dark-matter halo population in any direct or model
independent way.  For these reasons we restrict our attention to
quasars at redshifts beyond $z\sim3.7$, where we can relate the
evolution in the quasar luminosity function directly to evolution of
the Press-Schechter~(1974) mass function.

In \S~\ref{params} we introduce our parameterisation of SMBH/quasar
evolution, discuss the relation of these parameters to quasar
observables, and present our method of parameter estimation as well as
our choice of prior probabilities for the different variables. Our
approach is an extension of the formalism described in Wyithe \&
Padmanabhan~(2005, hereafter Paper-I). In \S\ref{constraints} and
\S\ref{lifetime} we describe the constraints that the available data
impose on physical and model parameters. The effects of scatter in the
relation between quasar luminosity and halo mass and of the choice of
parameterisation and prior probabilities are discussed in
\S\ref{scatter} and \S\ref{exp}. 
Finally we compare some simple physical models with our derived
constraints (\S\ref{compare}) before summarising our results in
\S\ref{conclusion}. Throughout the paper we adopt the set of
cosmological parameters determined by the {\em Wilkinson Microwave
Anisotropy Probe} (WMAP, Spergel et al. 2003), namely mass density
parameters of $\Omega_{m}=0.27$ in matter, $\Omega_{b}=0.044$ in
baryons, $\Omega_\Lambda=0.73$ in a cosmological constant, and a
Hubble constant of $H_0=71~{\rm km\,s^{-1}\,Mpc^{-1}}$. For the
primordial power-spectrum of density fluctuations, we adopt a
power-law slope $n = 1$, and the fitting formula to the exact transfer
function of Bardeen et al.~(1986).

\section{Model Parameters and Constraints}
\label{params}

\subsection{Parameters}

In this section we introduce our parameterisation of SMBH evolution.
Our aim is to describe a parameterised, rather than a physical, model
of SMBH evolution. We do not attempt to assign values for these
parameters in order to represent a particular physical theory. Rather
our goal is to consider all values so as to elucidate the acceptable
range of parameter values as well as any degeneracies. The
parameterisation chosen provides a general set of models within which
the various different possible physical models are contained.
Therefore by using this set of models to constrain the allowed values
for these parameters, we may determine which physical processes can be
unambiguously extracted from the available data.

We define halo mass $M$ to be the mass of a halo hosting a quasar with
a luminosity $\mathcal{M}_{1450}=-26.7$ at $z=4.8$.  The relation
between quasar luminosity $L$ and halo mass $M$ may have a mass
dependence and a redshift dependence in addition to a normalisation
($L_0/M_0$). We parameterise this dependence using two indices
$\delta$ and $\gamma$ as
\begin{equation}
\label{BHev}
\frac{L(z)}{L_0}=\left(\frac{M}{M_0}\right)^\delta(1+z)^\gamma.
\end{equation}

The Press-Schechter~(1974) mass function $n(M,z)$, [with the
modification of Sheth \& Tormen~(2002) that will be adopted throughout
our discussion] yields the number density $N(>M(z),z)$ of dark matter
halos above some mass $M(z)$ at redshift $z$. If luminous quasars
reside in a fraction $\epsilon$ of such dark-matter halos, then the
observed number density of quasars is given by the product of two
factors:
\begin{equation}
N(<\mathcal{M}_{1450}) = N(>M(z),z)\tau
\end{equation}
where
\begin{equation}
\tau \equiv \epsilon \ {\rm
min}\{t_{\rm q}/H^{-1}(z),1\},
\end{equation}
$t_{\rm q}$ is the (unknown) quasar lifetime and $H^{-1}(z)$ is the
Hubble time (see e.g., Estathiou \& Rees~1998).  Our fourth parameter
is introduced via the $z$ dependence of $t_q$. While $\epsilon$ could
also change with $z$ due to various effects (dust obscuration, beaming
angle etc.), we expect the dominant contribution to evolution in
$\tau$ to come from $t_{\rm q}$. Nevertheless we allow for growth of
$\epsilon$ with time.  The evolution of $\tau$ can be parameterised by
\begin{equation}
\label{tauev}
\frac{\tau(z)}{\tau_0}=(1+z)^\alpha.
\end{equation}
The parameters $\alpha$, $\gamma$ and $\delta$, in addition to
$\log_{10} M$ therefore govern the evolution of SMBHs. In this paper
our goal is to constrain the allowed values of these parameters using
observations of high redshift quasars. (Note that the model described
above assumes an arbitrarily chosen a power-law parameterisation. The
effect of modifying this choice is investigated in \S\ref{exp}.)

\subsection{Observational constraints}

There are 3 observational constraints which we will use to identify
the allowed range of parameters governing the evolution of SMBHs
($\alpha$, $\delta$ and $\gamma$) as well as halo mass $M$.  Let us
discuss key observations that constrain these.

First, observations of high redshift quasars (Fan et
al.~2001;2003;2004) reveal an exponential decline in the quasar
population with redshift of the form
\begin{equation}
\label{LF}
\Psi(\mathcal{M}_{1450}<-26.7,z)\propto 10^{B_{\rm obs}\times z},
\end{equation}
suggesting approximate constancy of $B$. As a measure of the rate at
which luminous quasars appear, we therefore use the exponential slope
$(B)$ of $\tau N(>M(z),z)$, which according to the above
parameterisation is defined as
\begin{eqnarray}
\nonumber
\label{PSslope}
B&=&0.434\frac{\alpha}{1+z} + \frac{\partial\log_{10} {N(>M,z)}}{\partial z}\\
&-& \frac{0.434}{(1+z)}\frac{\gamma}{\delta}\frac{d\log_{10} N(>M,z)}{d\log_{10} M} 
\end{eqnarray}

Second, observations of high redshift quasars (Fan et al.~2001) also
reveal a power-law decline in the density of quasars per unit
luminosity as a function of luminosity of the form
\begin{equation}
\label{LF2}
\frac{d\Psi}{dL}\propto L^\beta.
\end{equation}
As a measure of the relative numbers of quasars with different
luminosities at a single redshift we use the power-law slope of
$n(M,z)$
\begin{equation}
\label{beta}
\beta = \frac{1}{\delta}\frac{d\log_{10}(n)}{d\log_{10}(M)}.
\end{equation}
The factor of $1/\delta$ accounts for the nonlinear relation between
luminosity and halo mass.

Third, the ratio of SMBH to host dark-matter halo mass is determined
locally. Given a SMBH to host dark-matter halo mass ratio $R(M,z)$ at
high redshift, we can compute an extrapolated ratio at $z=0$ for halos
of mass $M_0=10^{12}M_\odot$ using the parameters $\delta$ and
$\gamma$.
\begin{eqnarray}
\label{ratio}
R_0\left(10^{12}M_\odot,0\right) = R\left(M,z\right)\left(\frac{M}{10^{12}M_\odot}\right)^{1-\delta}(1+z)^{-\gamma}
\end{eqnarray}
If we estimate the SMBH mass powering the high redshift quasars, then
we can calculate $R(M,z)$ for a given value of host mass $M$ and
compare the extrapolated value of $R_0$ for halos of $10^{12}M_\odot$
at $z=0$ with the observations.

In this paper we assign Gaussian probabilities for the observed
distributions of $B$, $\beta$ and $R_0$. In paper-I, based on the data
of Fan et al.~(2001;2003;2004) we found that $B_{\rm obs}$ is well
constrained with a mean $\bar{B}=-0.49$ and variance $\Delta
B=0.07$. Fan et al.~(2001) find $\bar{\beta}=-2.58$ and
$\Delta\beta=0.23$. Ferrarese~(2002) determines the ratio $M_{\rm
bh}/M$ for $M=10^{12}M_\odot$ and finds a significant dependence on
the assumed mass-profile, with values ranging between $\log_{10}
R_0=-5$ (for an NFW profile) and $\log_{10} R_0=-5.6$ (for an singular
isothermal profile). Since the true profile is likely to lie between
these extremes, we adopt a mean of $\log_{10}\bar{R_0}=-5.3$ and a
variance $\Delta \log_{10} R_0=0.3$.

\subsection{A-posteriori parameter estimation}

From Bayes theorem, the joint a-posteriori probability distribution
for the parameters of interest ($\alpha$, $\delta$, $\gamma$ and $M$)
as well as $\sigma_8$ is
\begin{eqnarray}
\label{d4P}
\nonumber
\left.\frac{d^5P}{d\alpha d\gamma d\delta d\log_{10} M d\sigma_8}\right|_{\rm obs} &\propto&\\
&&\hspace{-40mm} L_{B} L_\beta L_R \frac{dP_{\rm prior}}{d\gamma}
\frac{dP_{\rm prior}}{d\alpha}\frac{dP_{\rm
prior,obs}}{d\delta}\frac{dP_{\rm prior}}{d\log_{10}{M}}\frac{dP_{\rm
obs}}{d\sigma_8}.
\end{eqnarray}
In what follows we will consider both the prior and observed
constraints for $\delta$, hence the designation $P_{\rm prior,obs}$ in
equation~(\ref{d4P}). Since $\sigma_8$ has an observed distribution,
we may marginalise over its dependence
\begin{eqnarray}
\label{d3P}
\nonumber
\left.\frac{d^4P}{d\alpha d\gamma d\delta d\log_{10} M}\right|_{\rm obs} &&\propto \\
&&\hspace{-25mm}\int_0^\infty d\sigma_8 \frac{dP_{\rm
prior}}{d\sigma_8}\left.\frac{d^5P}{d\alpha d\gamma d\delta d\log_{10}
M d\sigma_8}\right|_{\rm obs}.
\end{eqnarray}
The likelihoods in equation~(\ref{d4P}) for $B$ and $\beta$ are
\begin{equation}
L_B = e^{\frac{-1}{2}(\frac{B-\bar{B}}{\Delta B})^2}\hspace{5mm}
\mbox{and}\hspace{5mm}L_\beta =
e^{\frac{-1}{2}(\frac{\beta-\bar{\beta}}{\Delta \beta})^2}.
\end{equation}

In what follows we will consider both the case in which the constraint
of the SMBH-halo mass ratio is not included, (so that $L_R=1$), and
the case in which the constraint of the SMBH-halo mass ratio is
included. In the later case (which requires an assumption about the
SMBH mass ($M_{\rm bh}$), and extrapolation of equation~(\ref{BHev})
between $z=0$ and $z\sim4.8$), we have
\begin{equation}
\label{LR}
L_R = \int_{-\infty}^\infty d\log_{10} M_{\rm bh} \frac{dP_{\rm prior}}{d\log_{10} M_{\rm bh}} e^{\frac{-1}{2}(\frac{R-\bar{R}}{\Delta R})^2},
\end{equation}
where we have marginalised over a prior dependence on SMBH mass
powering the high redshift quasars.

\subsection{Prior parameter distributions}
\label{priors}

In this section we specify the prior probability distributions
($P_{\rm prior}$) for parameters under investigation. In cases where
parameters are independently constrained we present our representation
of the observed distribution ($P_{\rm obs}$).

The value of $B$ is independent of the magnitude of the unknown quasar
lifetime. However in addition to the halo mass, the value of $B$
depends on the form of the redshift \emph{evolution} of $\tau$. The
$z$ dependence of $t_q$ is handled by using the parameter $\alpha_{\rm
lt}$.  The expected range of $\alpha_{\rm lt}$ may be set by the
following considerations. First, if the quasar lifetime is determined
by the mass $e$-fold timescale of the SMBHs, then $t_{\rm q}$ is
independent of redshift, $\tau\propto 1/H^{-1}(z)$ and $\alpha_{\rm
lt}\approx3/2$.  Second, if the quasar lifetime is determined by the
dynamical timescale at $z$, then $t_{\rm q}\approx H^{-1}(z)$ making
$\tau$ independent of redshift and $\alpha_{\rm lt}\simeq0$. (This is
also true if $t_{\rm q}> H^{-1}(z)$.) The evolution of the parameter
$\tau$ also depends on the evolution of
$\epsilon\propto(1+z)^{\alpha_\epsilon}$. We expect the value of
$\epsilon$ to grow with time, hence $\alpha_\epsilon<0$. Thus the
evolution of $\tau$ is parameterised according to
$\tau\propto(1+z)^\alpha$, where $\alpha=\alpha_{\rm
lt}+\alpha_\epsilon$.  We take the prior probability for $\alpha$ to
be flat over a wide range of values
\begin{eqnarray}
\nonumber
\frac{dP_{\rm prior}}{d\alpha}&=&\frac{1}{\alpha_{\rm max}-\alpha_{\rm min}}\\
&&\mbox{where}\hspace{3mm}\alpha_{\rm min}=-2,\alpha_{\rm max}=4.
\end{eqnarray}

In addition, the value of $B$ depends on the evolution of the relation
between luminosity and halo mass ($\gamma$). In paper-I we have shown
that $\gamma>0$ under the assumption of constant $\epsilon$, by
rejecting the null-hypothesis that $\gamma=0$. There are two
components of the parameter $\gamma$. First, the SMBH to halo mass
ratio can vary with redshift (a dependence described by a parameter
$\gamma_{\rm ratio}$). Second the accretion rate could vary with
redshift (a dependence which we describe by a parameter $\gamma_{\rm
acc}$). Since the accretion rate is close to Eddington at high
redshift (Fan et al.~2004), we expect $\gamma_{\rm acc}>0$. The
combined evolution of the quasar luminosity to halo mass ratio may
then be described by $\gamma=\gamma_{\rm ratio}+\gamma_{\rm acc}$.
Feedback scenarios at constant accretion rate imply $\gamma=2.5$. We
use the following broad prior-probability distribution for $\gamma$
which includes these values.
\begin{equation}
\frac{dP_{\rm prior}}{d\gamma} = \frac{1}{\gamma_{\rm max}-\gamma_{\rm min}} \hspace{3mm} \mbox{where} \hspace{3mm} \gamma_{\rm min}=-1,\gamma_{\rm max}=9/2
\end{equation}

The value of $\beta$ is independent of the unknown values of the
quasar lifetime (including its evolution $\alpha$) as well as
$\gamma$. However in addition to the halo mass, the value of $\beta$
depends on the form of the relation between quasar luminosity and halo
mass ($\delta$). As before there are two components to
$\delta$. First, the SMBH to halo mass ratio can vary with halo mass
at fixed redshift (a dependence described by a parameter $\delta_{\rm
ratio}$). Second the accretion rate at fixed redshift could vary with
halo mass (a dependence which we describe by a parameter $\delta_{\rm
acc}$). Since the accretion rate is close to Eddington in the most
luminous high redshift quasars (Fan et al.~2004), we expect
$\delta_{\rm acc}>0$. The combined evolution may then be described by
$\delta=\delta_{\rm ratio}+\delta_{\rm acc}$ Locally, the SMBH to halo
mass relation is steeper than linear ($\delta>1$), while feedback
driven scenarios at constant accretion rate, as well as observation
prefer $\delta\sim4/3-5/3$. We consider two cases for prior knowledge
of $\delta$. First, we use the following prior, which ignores
externally derived constraints
\begin{equation}
\label{deltaprior}
\frac{dP_{\rm prior}}{d\delta} = \frac{1}{\delta_{\rm max}-\delta_{\rm min}} \hspace{3mm} \mbox{where} \hspace{3mm} \delta_{\rm min}=1/2,\delta_{\rm max}=5/2. \\
\end{equation}
Second, we assume that the accretion rate is insensitive to halo mass
($\delta_{\rm acc}=0$), so that at fixed redshift the SMBH mass
depends only on the halo mass.  Locally, direct estimates of SMBH and
host mass can be made through observations of galaxy dynamics.  These
observations reveal a correlation between SMBH mass and the
characteristic velocity of the surrounding stellar spheroid
(e.g. Merritt \& Ferrarese~2001; Tremaine et al.~2002), and by
extension of the host dark matter halo (Ferrarese~2003). We therefore
also consider the case where $\delta$ is hypothesised to be redshift
independent so that the observed constraint can be applied at high
redshift. In this case we take
\begin{equation}
\label{deltaobs}
\frac{dP_{\rm obs}}{d\delta} = \frac{1}{\sqrt{2\pi}\Delta\delta_{\rm max}} e^{-\frac{1}{2}(\frac{\bar{\delta}-\delta}{\Delta\delta})^2},
\end{equation}
with a mean and variance of $\bar{\delta}=1.5$ and
$\Delta\delta=0.25$.

Comparison of the high redshift SMBH to halo mass ratio with the ratio
expected by extrapolating the local value to high redshift requires an
assumption about the SMBH mass. SMBHs powering the highest redshift
quasars are believed to have masses of $M_{\rm bh}\sim10^9M_\odot$
(Fan et al.~2001). We consider the following prior probability for the
SMBH mass ($M_{\rm bh}$) powering the high redshift quasars.
\begin{eqnarray}
\nonumber
\frac{dP_{\rm prior}}{d\log_{10} M_{\rm bh}} &=& \frac{1}{\log_{10} M_{\rm bh,max}-\log_{10} M_{\rm bh,min}}, \\
&&\hspace{-20mm}\mbox{where} \hspace{3mm} \log_{10} M_{\rm bh,min}=8.5,\log_{10} M_{\rm
bh,max}=9.5
\end{eqnarray}

Of course, all constraints depend on the value of $\log_{10} M$. SMBHs
powering the highest redshift quasars are though to have masses in
excess of $10^8M_\odot$, implying halo masses of
$M>10^9M_\odot$. Galaxies are not observed to have halos in excess of
$10^{14}M_\odot$. Hence we choose a prior probability for $\log_{10}
M$ of the form
\begin{eqnarray}
\nonumber
\frac{dP_{\rm prior}}{d\log_{10} M} &=& \frac{1}{\log_{10} M_{\rm max}-\log_{10} M_{\rm min}}, \\
&&\hspace{-10mm}\mbox{where} \hspace{3mm} \log_{10} M_{\rm min}=9,\log_{10} M_{\rm max}=14.
\end{eqnarray}

Finally, The evaluation of the PS mass-function, and hence of the
parameters $B$ and $\beta$ depend on the adopted cosmology, with the
greatest dependence being on $\sigma_8$. We utilise the observed
distribution for $\sigma_8$ from {\em WMAP} (Spergel et al.~2003)
\begin{equation}
\frac{dP_{\rm obs}}{d\sigma_8} = \frac{1}{\sqrt{2\pi}\Delta\sigma_8}e^{\frac{-1}{2}(\frac{\bar{\sigma}_8-\sigma_8}{\Delta\sigma_8})^2},
\end{equation}
with mean and variance of $\bar{\sigma_8}=0.84$ and
$\Delta\sigma_8=0.04$.

\section{Parameter Constraints}
\label{constraints}

In this section we present a-posteriori, joint probability
distributions for combinations of $\alpha$, $\gamma$, $\delta$ and
$\log_{10} M$. These distributions were obtained by marginalising
equation~(\ref{d3P}) over the other parameters. For example the joint
probability for $\gamma$ and $\log_{10}M$ may be found from
\begin{equation}
\frac{d^2P}{d\gamma d\log_{10} M} \propto \int_{-\infty}^\infty d\alpha \int_{-\infty}^\infty d\delta \frac{d^4P}{d\alpha d\gamma d\delta d\log_{10} M}.
\end{equation}
We also determine a-posteriori cumulative probability distributions
for individual parameters. For example, the cumulative probability for
$\log_{10}M$ is
\begin{eqnarray}
\nonumber
P(<\log_{10} M) &\propto&\int_{-\infty}^{\log_{10} M} d\log_{10} M^\prime  \\
&&\hspace{-15mm}\int_{-\infty}^\infty d\alpha \int_{-\infty}^\infty d\delta\int_{-\infty}^\infty d\gamma \frac{d^4P}{d\alpha d\gamma d\delta d\log_{10} M^\prime}.
\end{eqnarray}

We show results in 3 different cases:

\begin{itemize}

\item Case-I: We begin by determining the constraints that can be made
considering only the limits on $B$ and $\beta$ derived from the SDSS
quasars, i.e. $L_R=1$ and assumption only of the prior probability
(equation~\ref{deltaprior}) for $\delta$. In this case we are trying
to determine 4 parameters from two constraints, and so are strongly
under-constrained.

\item Case-II: We next determine the constraints that can be made
considering the limits on $R_0$ in addition to those on $B$ and
$\beta$, thus $L_R$ is given by equation~(\ref{LR}). As in Case-I, we
assume a flat prior probability (equation~\ref{deltaprior}) for
$\delta$. The addition of the third constraint partially breaks the
degeneracies in Case-I.

\item Case-III: Finally we replace the prior probability
(equation~\ref{deltaprior}) for $\delta$, with the distribution
observed locally (equation~\ref{deltaobs}).

\end{itemize}

\begin{figure*}
\vspace*{123mm}
\includegraphics{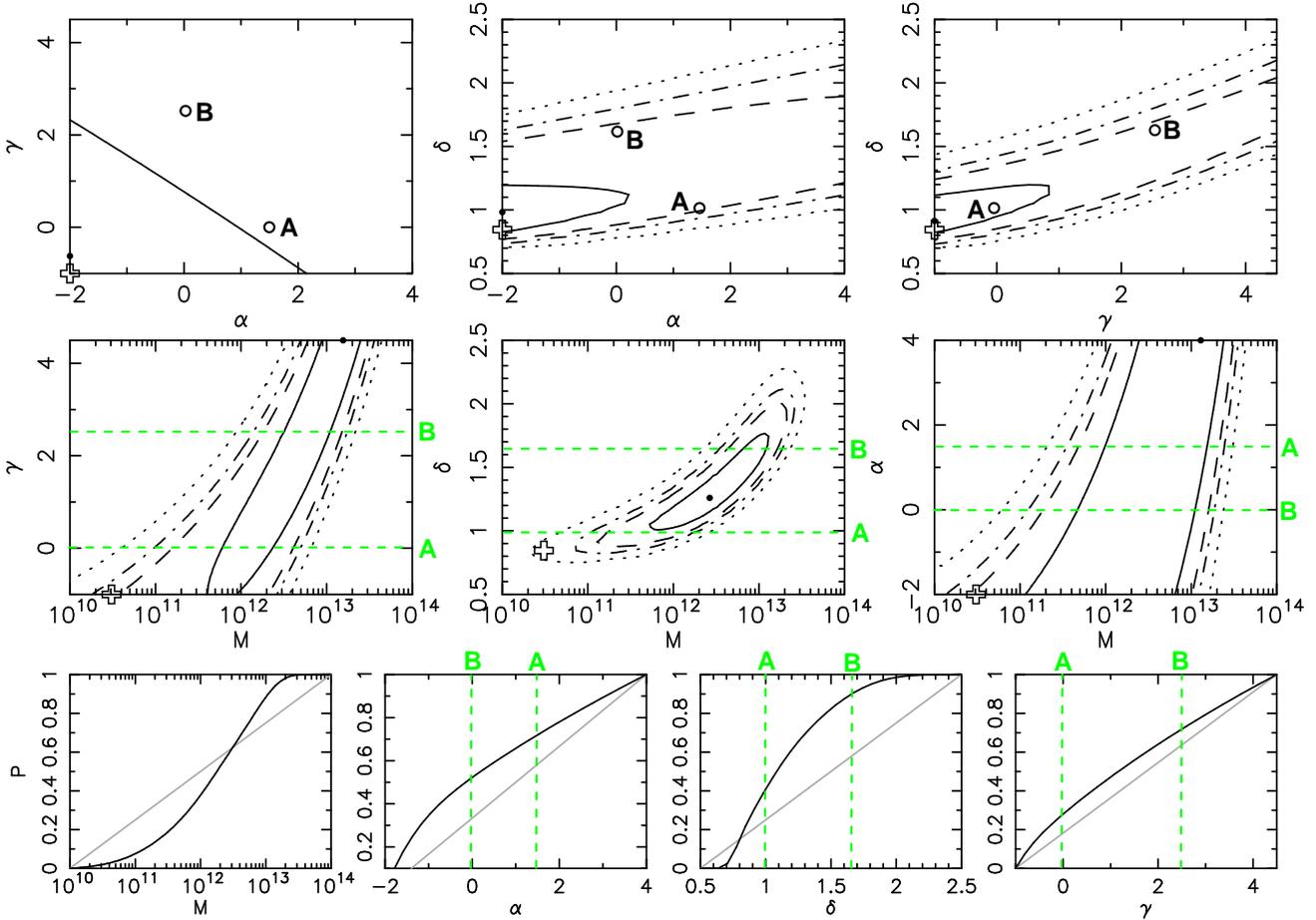}
\caption{Constraints on power-law parameters, Case-I: The various panels in the upper two rows show contours for the marginalised a-posteriori joint probability distributions for combinations of the parameters $\gamma$, $\delta$, $\alpha$ and $M$. The dots show the location of the maximum likeli-hood within the marginalised distributions, while the cross shows the location of the maximum likeli-hood within the full 4-dimensional parameter space. The lower row shows marginalised a-posteriori cumulative probability distributions for $\gamma$, $\delta$, $\alpha$ and $M$ (dark lines). The light lines show the respective prior probability distributions. The large open circles in the upper panels, and the light dashed lines in the central and lower panels show the location in parameter space of models A and B described in \S~\ref{compare}.}
\label{fig1} 
\end{figure*}

\begin{figure*}
\vspace*{123mm}
\includegraphics{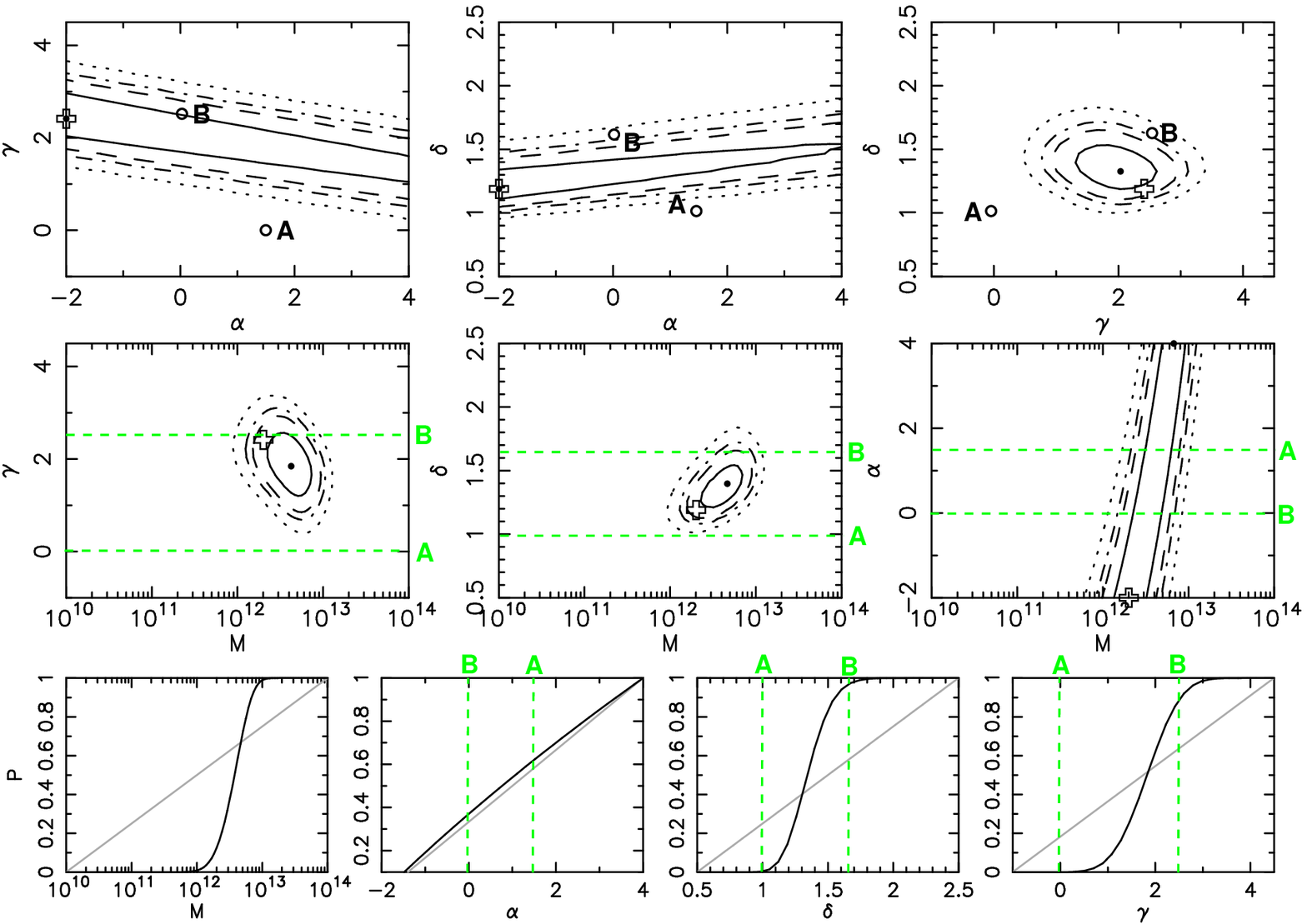}
\caption{Constraints on power-law parameters, Case-II: The various panels in the upper two rows show contours for the marginalised a-posteriori joint probability distributions for combinations of the parameters $\gamma$, $\delta$, $\alpha$ and $M$. The dots show the location of the maximum likeli-hood within the marginalised distributions, while the cross shows the location of the maximum likeli-hood within the full 4-dimensional parameter space. The lower row shows marginalised a-posteriori cumulative probability distributions for $\gamma$, $\delta$, $\alpha$ and $M$ (dark lines). The light lines show the respective prior probability distributions. The large open circles in the upper panels, and the light dashed lines in the central and lower panels show the location in parameter space of models A and B described in \S~\ref{compare}.}
\label{fig2} 
\end{figure*}

\begin{figure*}
\vspace*{195mm}
\includegraphics{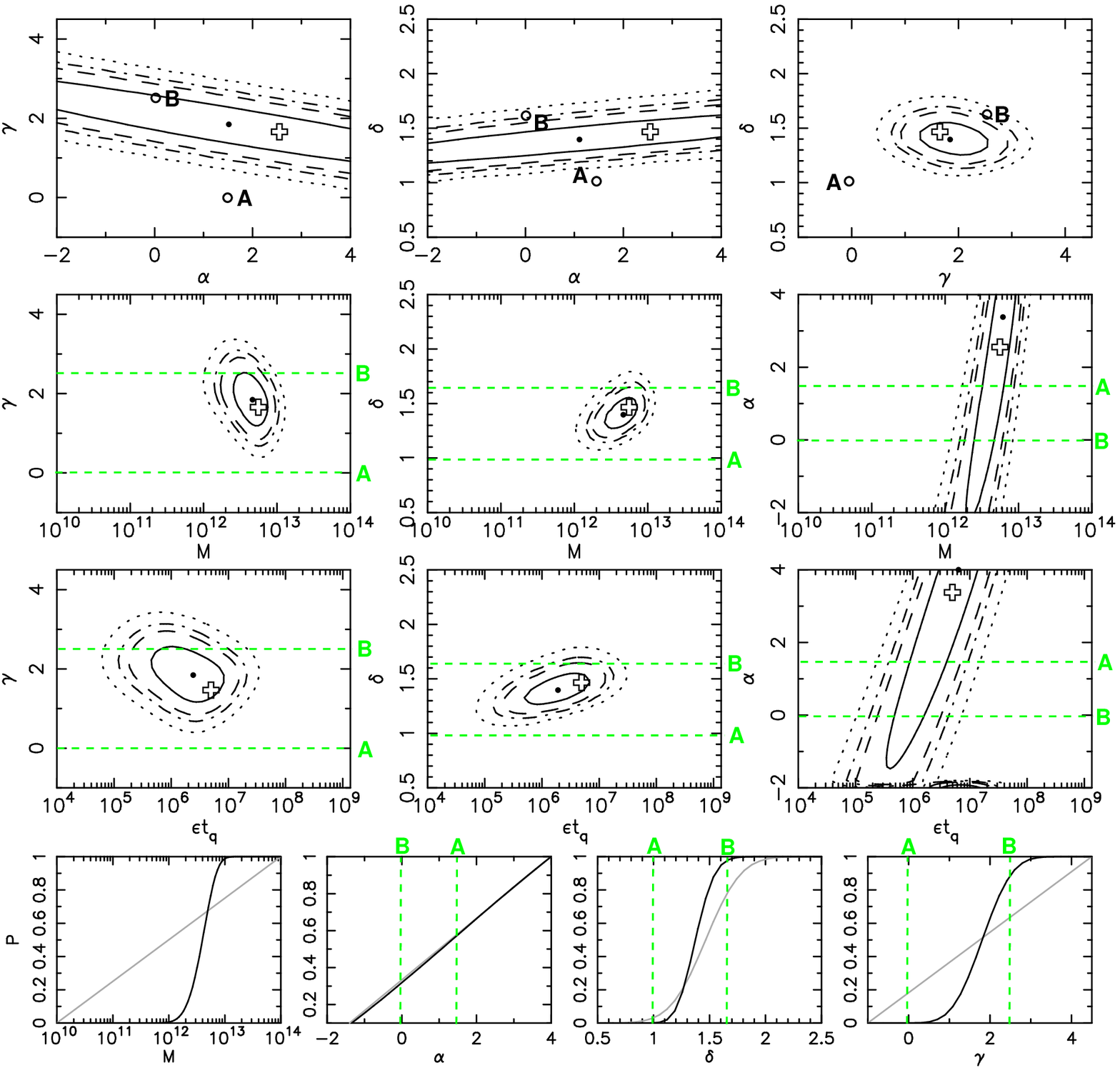}
\caption{Constraints on power-law parameters, Case-III: The upper and second rows show contours for the marginalised a-posteriori joint probability distributions for combinations of the parameters $\gamma$, $\delta$, $\alpha$ and $M$. The third row shows distributions with $M$ replaced by $(\epsilon t_{\rm lt})$. The dots show the location of the maximum likeli-hood within the marginalised distributions, while the cross shows the location of the maximum likeli-hood within the full 4-dimensional parameter space. The lower row shows marginalised a-posteriori cumulative probability distributions for $\gamma$, $\delta$, $\alpha$ and $M$ (dark lines). The light lines show the respective prior probability distributions. The large open circles in the upper panels, and the light dashed lines in the panels of the second, third and lower rows show the location in parameter space of models A and B described in \S~\ref{compare}.}
\label{fig3} 
\end{figure*}

Results are presented in Figures~\ref{fig1}-\ref{fig3} for
Cases-I--III respectively. The sets of panels in the upper two rows
show contours of $\left.d^2P/(d\alpha d\gamma)\right|_{\rm obs}$,
$\left.d^2P/(d\alpha d\delta)\right|_{\rm obs}$, $\left.d^2P/(d\delta
d\gamma )\right|_{\rm obs}$, $\left.d^2P/(d\gamma d\log_{10}
M)\right|_{\rm obs}$, $\left.d^2P/(d\delta d\log_{10} M)\right|_{\rm
obs}$, and $\left.d^2P/(d\alpha d\log_{10} M)\right|_{\rm obs}$. For
the display of joint probability distributions, the solid, dashed,
dot-dashed and dotted contours correspond to values that are 0.61,
0.26, 0.14 and 0.036 times the distributions peak value, whose
position is marked by a dot. When applied to a Gaussian distribution,
these values correspond to the 1, 2, 3 and 5-sigma levels
respectively. We also show a-posteriori cumulative probability
distributions for individual parameters in the lower rows of
Figures~\ref{fig1}-\ref{fig3} (dark lines) together with the
corresponding prior probability distributions (light lines).

For each case we determine the values corresponding to the maximum
likeli-hood solution normalised with respect to a maximum possible
value of unity. Maximum likeli-hoods of order unity imply that the
model is able to accommodate all constraints simultaneously. The
values corresponding to the maximum likeli-hood solution are shown in
Figures~\ref{fig1}-\ref{fig3} (large crosses). For Case-I the maximum
value of the likelihood is given by $L_{I}=L_\beta L_{B}$. For Case-II
we include the constraint of local mass ratio so that the likeli-hood
becomes $L_{II}=L_\beta L_{B}L_R$. Finally the addition of prior
information on the parameter $\delta$ leads to a maximum value for the
likelihood in Case-III of $L_{III}=L_\beta
L_{B}L_Re^{-\delta^2/(2\Delta\delta)}$.

We consider each case in turn:

{\em Case-I:} In the first case we determine what can be learned
regarding the connection between SMBHs and host halos, using only the
evolution in the number counts of quasars at high redshift and the
slope of the high redshift quasar luminosity function. The results are
shown in Figure~\ref{fig1}. In this case we have two parameters that
may be measured ($\beta$ and $B$) but 4 model parameters that lack
prior constraints. Clearly this problem is
under-constrained. Inspection of equations~(\ref{PSslope}) and
(\ref{beta}) shows that $\beta$ depends only on $M$ and $\delta$,
while $B$ is a function of all four parameters. Thus we would expect a
degeneracy in the determination of parameters $\delta$ and $M$ which
is driven by the observed luminosity function slope $\beta$. This
degeneracy is clearly seen in the center panel of Figure~\ref{fig1},
although it is broken at high and at low $M$ by the additional
constraints from $B$. Thus even in this under-constrained case we are
able to obtain an estimate of mass, and show that $\delta>1$. The
preferred value for the parameter $\delta$ is quite insensitive to the
value of either $\alpha$ or $\gamma$. Because the measured value of
$B$ is degenerate between $M$ (or $\delta$ through the constraint of
$\beta)$, $\gamma$ and $\alpha$, there will be a degeneracy along a
line in the three dimensional parameter space. As a result we see that
no constraints are possible in the $\gamma-\alpha$ plane, while there
are strong degeneracies between $\alpha$ and $M$ and between $\gamma$
and $M$.  The maximum value of the likelihood is $L_{I}=0.92$,
implying that the best fit model is a very good fit to the high
redshift data. The set of best fit parameters is
$(\log_{10}M,\gamma,\delta,\alpha)=(10.48,-1,0.84,-2)$. The best fit
value of $\alpha$ is on the edge of the range of parameter space,
which is indicative of the degeneracies between $\alpha$ and other
parameters.

Parameters for which the data are the restrictive have a-posteriori
probability distributions that differ from the assumed
prior-probability distributions. As a result the lack of restriction
imposed on models for SMBH growth by the high redshift luminosity
function may be clearly seen in the lower row of Figure~\ref{fig1}.
While the high redshift luminosity function places limits on each of
the parameters $\delta$ and $M$, no limits may be placed on the
parameters $\alpha$ or $\gamma$. Additional observables must therefore
be included in order to constrain the models. We next turn to
inclusion of the likelihood based on the local SMBH to halo mass
ratio.

{\em Case-II:} We can break part of the degeneracy seen in Case-I by
including the likeli-hood for the local SMBH to halo mass ratio
(equation~\ref{LR}) as an additional constraint. This inclusion has a
two-fold effect on the parameter fit. Firstly, it couples the
dependence of the overall likeli-hood of a solution on $\delta$ and
$\gamma$ in a direct way. The likelihood for the ratio $R$ will
therefore be degenerate between $\delta$ and $\gamma$. Secondly, the
value of $\gamma$ is now constrained by the likelihood for $R$ in
addition to the likelihood for $B$. This breaks the three-way
degeneracy between $M$, $\alpha$ and $\gamma$. The interplay between
the different likelihoods and parameters is now coupled so that simple
interpretation of the solution is more difficult than in
Case-I. However we now have 3 observables ($B$, $\beta$ and $R$), from
which to constrain 4 parameters. The solution must therefore still be
subject to a certain degree of degeneracy.

The results are shown in Figure~\ref{fig2}. We find that the
combinations of $\gamma$ with $M$, $\delta$ with $M$ and $\gamma$ with
$\delta$ are well constrained. In particular we find $\gamma>0$. This
may also be seen by inspecting the a-posteriori cumulative probability
distributions. We see that $\gamma$, $\delta$ and $\log_{10}M$ have
a-posteriori distributions that differ substantially from their prior
probabilities, indicating that the data is able to constrain their
values. The value of $\delta$ lies between 1.1 and 1.6 (90\%), and is
larger than unity in agreement with the locally observed prior
probability distribution for $\delta$ that will be included in
Case-III. The mass is constrained to be larger than $10^{12.2}M_\odot$
and smaller than $10^{12.8}M_\odot$ (90\%), while $1<\gamma<2.7$
(90\%). Unless the accretion rate is changing rapidly with redshift
near $z\sim4.5$, the positive value of $\gamma$ implies a SMBH to halo
mass ratio that decreases with time. The main degeneracies due to the
remaining degree of freedom are between $\alpha$ and $M$ (lower
right), between $\alpha$ and $\delta$ (upper center) and between
$\alpha$ and $\gamma$ (upper left), although the dependencies on
$\alpha$ are weak. The maximum value of the likelihood is
$L_{II}=0.48$, implying that the best fit model remains a very good
fit to the data following the addition of the constraint on the mass
ratio. The set of best fit parameters is
$(\log_{10}M,\gamma,\delta,\alpha)=(12.3,2.41,1.19,-2.)$. The value of
$\alpha$ remains at the edge of the range of parameter space,
indicating that the degeneracies between $\alpha$ and other parameters
is not removed through the inclusion of the mass ratio.

{\em Case-III:} Finally we include a prior probability for $\delta$
(equation~\ref{deltaobs}) based on observations of SMBHs at low
redshift. The results are plotted in the upper and central rows of
figure~\ref{fig3}.  Strictly speaking the solution is still
under-constrained since there are 4 parameters and three
constraints. However the observed prior distribution for $\delta$ is
sufficiently narrow that the hypothesis of the slope in the local
relation ($\delta$) also holding at high redshift strongly constrains
the solution. The result is that both $\gamma$ and $M$ are tightly
constrained by the data. The maximum likeli-hood value is
$L_{III}=0.40$, with a set of best fit parameters
$(\log_{10}M,\gamma,\delta,\alpha)=(12.7,1.65,1.47,2.55)$. From the
lower row of Figure~\ref{fig3} we see that only $\gamma$ and
$\log_{10}M$ have a-posteriori distributions that differ substantially
from their prior probabilities. The data does not restrict $\delta$
much beyond the assumed prior probability in this case. However this
indicates that the preferred value of $\delta$ at $z\sim4.5$ is
similar to the local value as hypothesised. As in Case-II the data
shows a definite preferred solution for these quantities. SMBHs form a
smaller fraction of their hosts mass at early times than they do
today, $1<\gamma<2.8$ (90\%), with high redshift host masses of
$10^{12.2-12.8}M_\odot$ (90\%). The parameter $\alpha$ remains
unconstrained.

\begin{figure*}
\vspace*{90mm}
\includegraphics{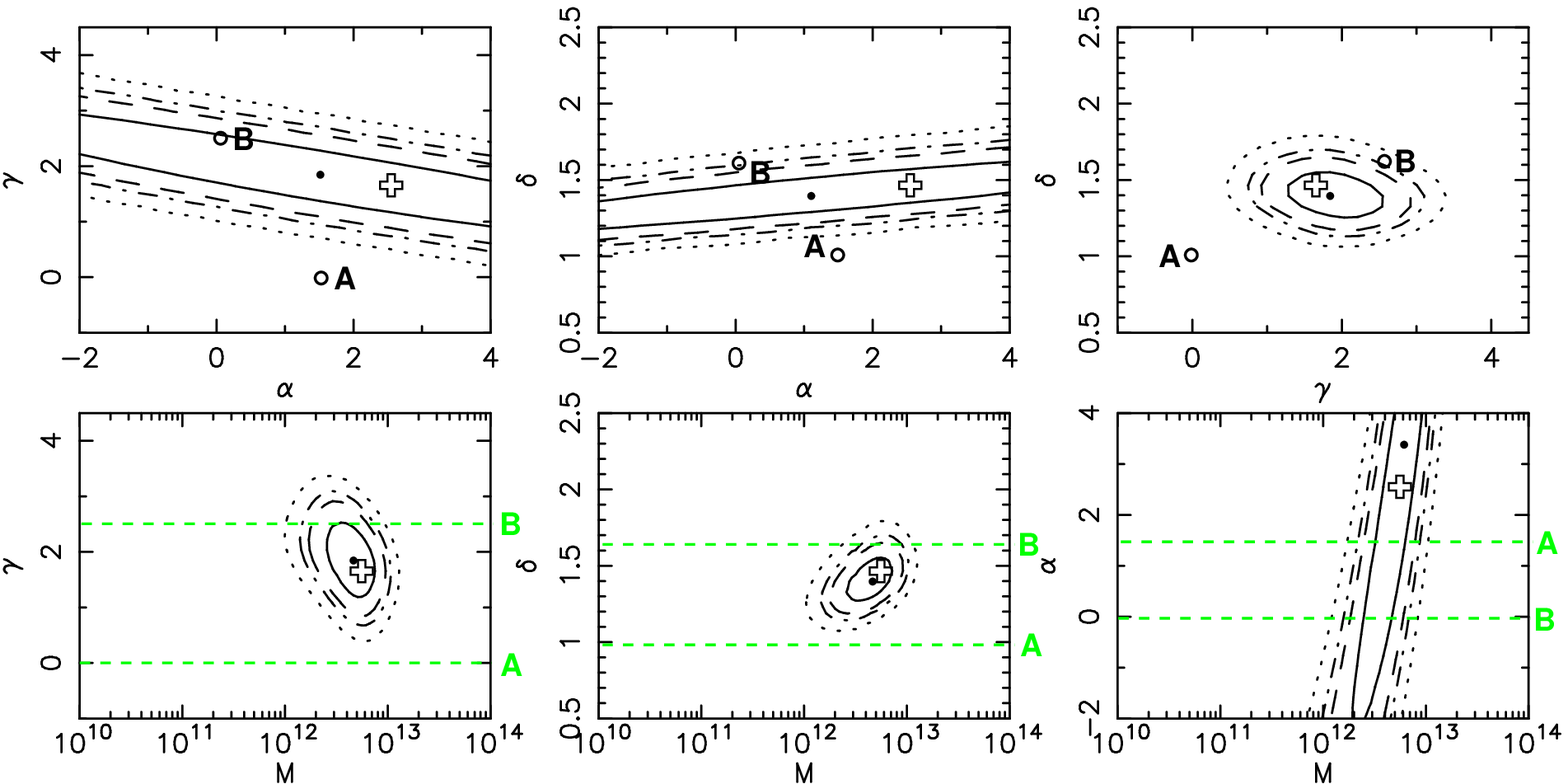}
\caption{Constraints on power-law parameters where scatter in the SMBH to halo mass ratio is included, Case-III: The various panels show contours for the marginalised a-posteriori joint probability distributions for combinations of the parameters $\gamma$, $\delta$, $\alpha$ and $M$. The dots show the location of the maximum likeli-hood within the marginalised distributions, while the cross shows the location of the maximum likeli-hood within the full 4-dimensional parameter space. The large open circles in the upper panels, and the light dashed lines in the lower panels show the location in parameter space of models A and B described in \S~\ref{compare}.}
\label{fig4} 
\end{figure*}

\section{quasar lifetime}
\label{lifetime}

If a fraction $\epsilon$ of dark matter halos contain SMBHs, then the
total lifetime of the quasar can be estimated (Martini \&
Weinberg~2001; Haiman \& Hui~2001) by dividing the quasar density
$\Psi(\mathcal{M}_{1450}<-26.7,z)$ by $\epsilon$ times the density $N(>M)$ of
halos larger than $M$, and then multiplying by the Hubble time (for
$t_{\rm q} < H^{-1}$), hence
\begin{equation}
\epsilon t_{\rm q}(\Psi,M) = H^{-1}(z) \frac{\Psi(z)}{N(>M,z)}
\end{equation}
There is therefore a one to one correspondence between $\epsilon
t_{\rm q}$ and $M$. As a result the joint a-posteriori distribution,
with $M$ replaced by $\epsilon t_{\rm q}$ may be obtained as
\begin{eqnarray}
\label{d3Ptq}
\nonumber
\left.\frac{d^4P}{d\alpha d\gamma d\delta d\log_{10} (\epsilon t_{\rm q})}\right|_{\rm obs} &\propto& \\
&&\hspace{-40mm}\int d\sigma_8 \frac{dP_{\rm prior}}{d\sigma_8}\left.\frac{d^5P}{d\alpha d\gamma d\delta d\log_{10} M d\sigma_8}\right|_{\rm obs} \frac{d\log_{10} M}{d\log_{10} (\epsilon t_{\rm q})},
\end{eqnarray}
where
\begin{equation}
\frac{d\log_{10} M}{d\log_{10} (\epsilon t_{\rm q})} = \left[\frac{d}{d\log_{10} M}\int d\Psi \frac{dP}{d\Psi}\epsilon t_{\rm q}(\Psi,M)\right]^{-1}.
\end{equation}

In the third row of figure~\ref{fig3} we show the joint probability
distributions (computed using equation~\ref{d3Ptq}) for $\alpha$ and
$\log_{10} (\epsilon t_{\rm q})$, $\gamma$ and $\log_{10} (\epsilon
t_{\rm q})$, and for $\delta$ and $\log_{10} (\epsilon t_{\rm
q})$. The distributions were computed using Case-III constraints. The
behavior is similar to that for the distributions of $\log_{10}M$ as
expected. If the occupation fraction is of order unity, then we find
lifetimes of $10^{6-7}$ years, in agreement with estimates at lower
redshifts. Observed quasar lifetimes therefore suggest an occupation
fraction $\epsilon$ that is of order unity. Since the Hubble time is
$\sim10^9$ years at $z\sim4.8$, we find that the occupation fraction
must be larger than $\epsilon\sim10^{-3}$--$10^{-2}$. 

\section{scatter in the luminosity--halo mass relation}
\label{scatter}

Up till now we have assumed a one-to-one correspondence between halo
mass and quasar luminosity (equation~\ref{BHev}). However we would
expect a scatter in this relation, due to (at least because of)
scatter in the $M_{\rm bh}-M$ relation, and due to variation in the
accretion rate. To assess the importance of this scatter, we make the
following modification to the calculation already described. We have
associated a quasar of luminosity $L$ with a halo of mass $M$. We can
regard the mass $M$ as characteristic, and replace the density of
halos $n(M,z)$ with an effective density
\begin{equation}
n_{\rm eff}(M,z) = \frac{1}{\sqrt{2\pi}\Delta m}\int_0^\infty dm' n(M',z) e^{\frac{-1}{2}(\frac{m'-m}{\Delta m})^2},
\end{equation}
which is obtained as a weighted average of the densities of halos that
house a quasar of luminosity $L$. Here $m\equiv\log_{10}M$. Effective
values of $B$, $\gamma$ etc. can then be calculated using $n_{\rm
eff}(M,z)$. We have computed the resulting probability distributions
for $\alpha$, $\gamma$, $\delta$ and $\log_{10} M$ as before in order
to assess the importance of scatter in
equation~(\ref{BHev}). Distributions corresponding to Case-III were
computed taking $\Delta m=0.5$ (so that 68\% of halos are contained
within 1 decade of mass), and are plotted in figure~\ref{fig4}.

The results are similar to those obtained in the absence of
scatter. In particular the contours for distributions move by less
than 1-sigma, and have quantitatively similar values.  The maximum
value of the likelihood is $L_{III}=0.38$, implying that the best fit
model remains a very good fit to the data following the addition of
scatter in the SMBH--halo mass relation. The set of best fit
parameters is
$(\log_{10}M,\gamma,\delta,\alpha)=(12.62,1.84,1.47,3.59)$.

\section{Dependence of results on choice of priors and parameterisation}
\label{exp}

In this paper our aim is to determine what physical properties may be
inferred from the high redshift quasar luminosity function within a
general, rather than a specific framework.  Two issues arise regarding
the statistical robustness of our results. Firstly, having chosen a
particular parameterisation for our model we then calculate
likelihoods for parameters using that parameterisation (for example
$\log_{10}M$, $\gamma$ and $\alpha$ in our prescription). We have
chosen flat prior probability distributions for these parameters and
hence find aposteriori probabilities for their values. Now if the
observations are able to genuinely constrain the values of these
parameters, then our results should be independent of the prior used
(within reason). We would therefore like to know whether our results
are sensitive to the choice of prior. Secondly, we have chosen a
powerlaw parameterisation for the redshift dependencies of the
lifetime and SMBH to halo mass ratio. These parameterisations may or
not provide an adequate description of reality. We would therefore
like to know whether the choice of parameterisation affects our
physical conclusions regarding the mass of host dark matter halos, or the
evolution of the SMBH to halo mass ratio. We discuss these issues in
the remainder of this section.

\subsection{Dependence on prior probabilities}

\begin{figure*}
\vspace*{90mm}
\includegraphics{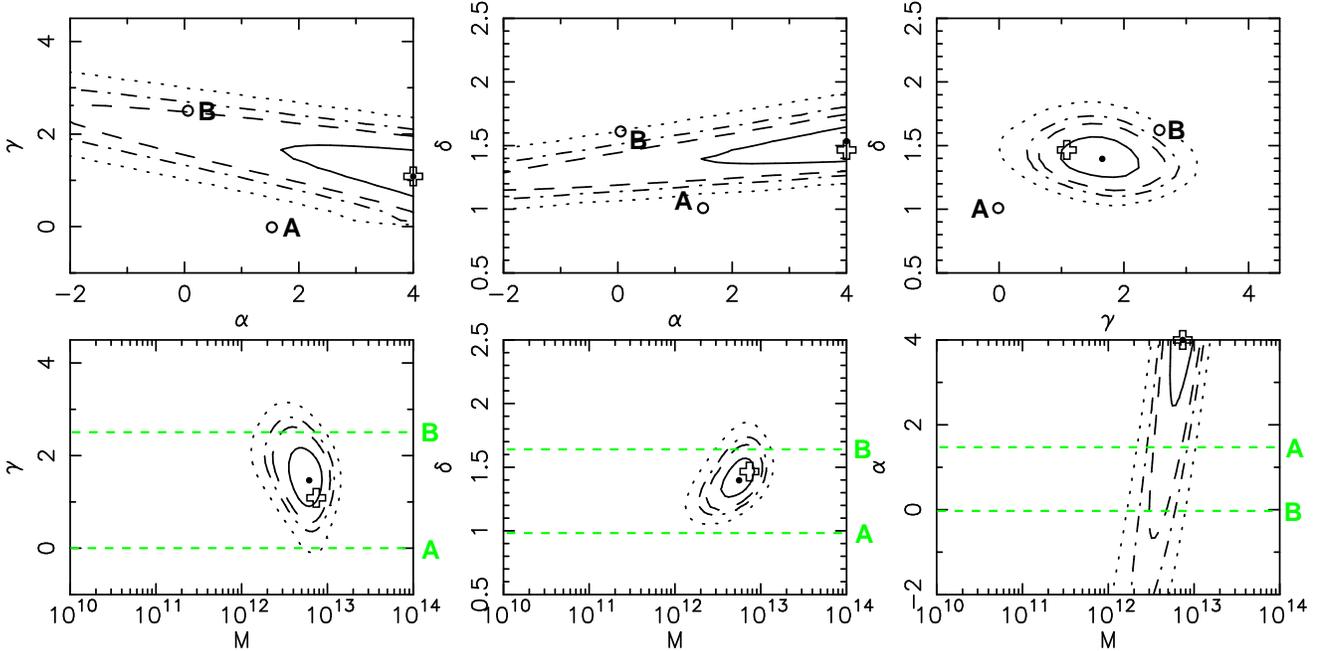}
\caption{Constraints on power-law parameters assuming alternative priors, Case-II: The various panels show contours for the marginalised a-posteriori joint probability distributions for combinations of the parameters $\gamma$, $\delta$, $\alpha$ and $M$. The dots show the location of the maximum likeli-hood within the marginalised distributions, while the cross shows the location of the maximum likeli-hood within the full 4-dimensional parameter space.}
\label{fig5} 
\end{figure*}

Up till now we have assumed prior probability distributions for
dimensionless parameters ($\alpha$, $\gamma$ and $\delta$) that are
flat (i.e. $dP_{\rm prior}/d\alpha\propto1$ etc). For the dimensioned
quantity ($M$) we have assumed a prior probability that is flat in the
logarithm [i.e. $dP_{\rm prior}/d\log_{10}(M)\propto1$]. While these
are the natural choices, we have repeated the calculations in Case-II
assuming priors for $\gamma$ and $\delta$ that are flat in the
logarithm (for $0<\gamma<9/2$ and $1/2<\delta<5/2$), and a prior
probability $dP_{\rm prior}/dM\propto1$, where
$10^9M_\odot<M<10^{14}M_\odot$ for host halo mass. We have not changed
the prior of the unconstrained parameter $\alpha$.

Figure~\ref{fig5} shows the resulting parameter constraints. The set
of best fit parameters is
$(\log_{10}M,\gamma,\delta,\alpha)=(12.87,1.08,1.47,4.0)$. The
constrained value for $M$ is quite similar to the original case
(figure~\ref{fig2}), thus the estimate of this (physical) parameter
would appear to be insensitive to the choice of prior. The values of
$\delta$ and $\gamma$ also take similar values indicating that they
are constrained by the data rather than by the prior. The conclusion that
the SMBH to halo mass ratio was larger in the past ($\gamma>0$) is
therefore not sensitive to the choice of prior. The value of $\alpha$
remains the least constrained, although positive values are preferred
in this case.

\subsection{Exponential parameter constraints}

\begin{figure*}
\vspace*{90mm}
\includegraphics{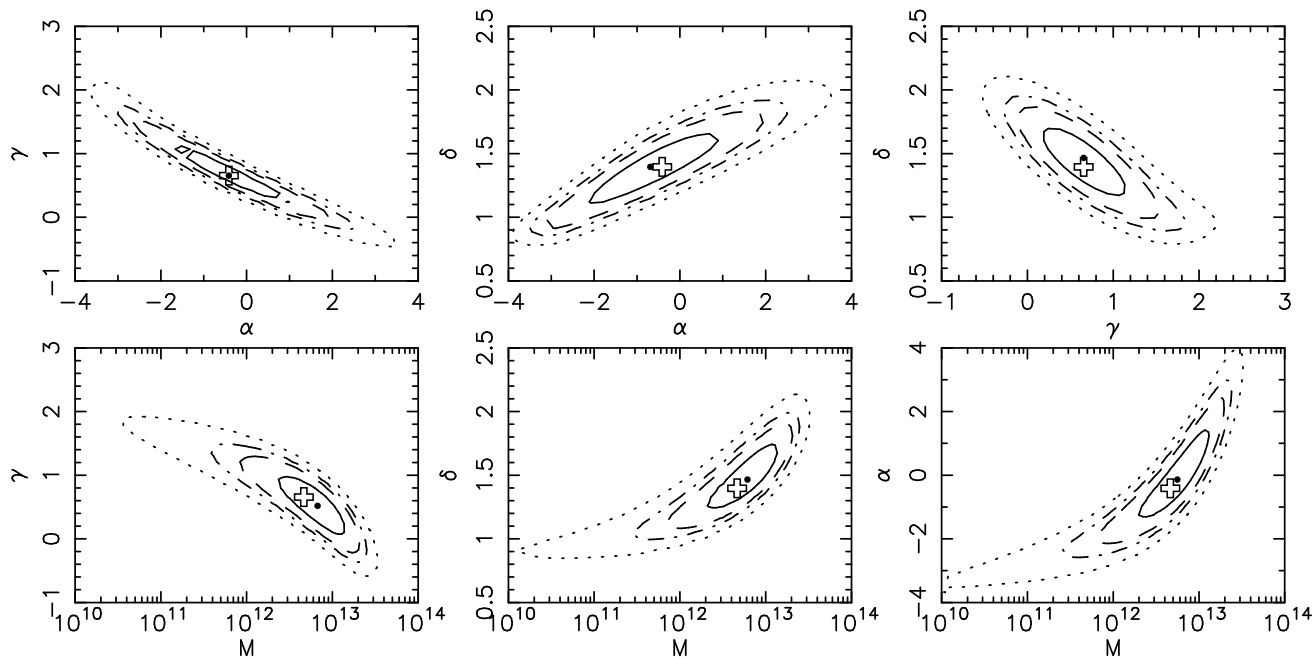}
\caption{Constraints on power-law ($\delta$) and exponential ($\alpha$ and $\gamma$) parameters, Case-III: The various panels show contours for the marginalised a-posteriori joint probability distributions for combinations of the parameters $\gamma$, $\delta$, $\alpha$ and $M$. The dots show the location of the maximum likeli-hood within the marginalised distributions, while the cross shows the location of the maximum likeli-hood within the full 4-dimensional parameter space.}
\label{fig6} 
\end{figure*}

The parameters $\alpha$ and $\gamma$ describe redshift dependencies of
the lifetime and SMBH to halo mass ratio respectively. Up till now we
have specified power-law evolutions for these quantities, and
constrained the allowed values of the corresponding parameters. Of
course it is only our theoretical prejudice that leads us to specify a
power-law, since we have no observational evidence for this form of
parameterisation. The assumption of a particular prior probability
distribution for a parameter (Bayes Postulate) leads to prior
probabilities for the physical parameter of interest (like the
life-time or mass ratio) that depend on the parameterisation
employed. We would therefore like to check that our choice of model
does not qualitatively bias the results. To check the effect of our
choice of parameterisation, we have re-calculated our results assuming
exponential rather than power-law evolution for those quantities that
vary with redshift (i.e. $\tau$ and $L(z)/L_0$). However we retain our
power-law form for mass ratio at fixed redshift ($\delta$) since this
is an observed rather than a postulated parameterisation.

Thus equations~(\ref{BHev}) and (\ref{tauev}) become
\begin{equation}
\label{BHev_exp}
\frac{L(z)}{L_0}=\frac{M_{\rm bh}(z)}{M_{\rm bh,0}}=\left(\frac{M}{M_0}\right)^\delta e^{\gamma z},
\end{equation}
and 
\begin{equation}
\label{tauev_exp}
\frac{\tau(z)}{\tau_0}=e^{\alpha z}, 
\end{equation}
which leads to evolution described by the equation 
\begin{eqnarray}
\label{PSslope_exp}
\nonumber
B&=&0.434\alpha + \frac{\partial\log_{10} {N(>M,z)}}{\partial z}\\
&+& 0.434\gamma\frac{d\log_{10} N(>M,z)}{d\log_{10} M}. 
\end{eqnarray}
The constraint from the local mass ratio (equation~\ref{ratio}) becomes 
\begin{equation}
\label{ratio_exp}
R_0\left(10^{12}M_\odot,0\right) = R\left(M,z\right)\left(\frac{M}{10^{12}M_\odot}\right)^{1-\delta}e^{-\gamma z},
\end{equation}
while the constraint based on the slope of the luminosity function
(equation~\ref{beta}) remains unchanged. Note that the relation of $B$
to $R$ as a function of $\gamma$ is altered by the new definition.

Figure~\ref{fig6} shows the resulting parameter constraints assuming
the analysis described in Case-III. The constrained value for $M$ is
quite similar to the power-law case (figure~\ref{fig3}), thus the
estimate of this (physical) parameter would appear to be insensitive
to the choice of parameterisation (assuming a monotonic evolution over
the range of interest). The maximum value of the likelihood is
$L_{III}=0.43$, implying that the best fit exponential model also
provides a very good fit to the data. The set of best fit parameters
is $(\log_{10}M,\gamma,\delta,\alpha)=(12.7,0.66,1.4,-0.41)$. The
values of $\alpha$ and $\gamma$ will clearly differ from the power-law
case. However the conclusion that the SMBH to halo mass ratio was
larger in the past ($\gamma>0$) holds in the exponential case as in
the power-law case. The exponential parameterisation allows a
preferred value of $\alpha\sim0$, indicating an occupation fraction
and/or duty-cycle that does not vary much with redshift.

\section{Comparison with simple physical models}
\label{compare}

Before concluding, we apply our idea to a couple of simple test case
models. Among the successful models of the high redshift quasar
luminosity function, two examples employ particularly simple but
different physical hypotheses. In the first, the quasar lifetime is
constant (perhaps set by the Salpeter time) with an occupation
fraction of unity, the SMBH makes up a constant fraction of the host
halo mass, and the quasar shines at its Eddington rate (Haehnelt,
Natarajan \& Rees~1998; Haiman \& Loeb~1998). We refer to this as
model-A. In the second model, the SMBH mass is set by feedback
(through energy conservation) over the dynamical time of the gas
reservoir, the occupation fraction is of order unity and the quasar
shines at its Eddington rate (Silk \& Rees~1998; Haehnelt, Natarajan
\& Rees~1998; Wyithe \& Loeb~2003). We refer to this as model-B. In
the present context these two models may be described by the parameter
sets ($\alpha_{\rm A}$,$\gamma_{\rm A}$,$\delta_{\rm A}$)=(1.5,0,1)
and ($\alpha_{\rm B}$,$\gamma_{\rm B}$,$\delta_{\rm B}$)=(0,2.5,1.66)
respectively. The locations of these models in parameter space are
shown in Figures~\ref{fig1}-\ref{fig5} by the the light dashed lines,
and the letters {\bf A} and {\bf B}.

Considering first the constraints in Case-I (Figure~\ref{fig1}), we
see that the degeneracies prohibit any conclusion regarding $\alpha$
or $\gamma$, while the preferred $\delta$ lies between the two
models. Indeed the two models may not be distinguished at all based on
the luminosity function data alone. On the other hand the simple
models described should also be able to accommodate the constraint
from the local SMBH to halo mass ratio.  In Case-II
(Figures~\ref{fig2} and \ref{fig5}) the different values of $\alpha$
for the two models are again indistinguishable. However in contrast
the values of $\gamma$ and $\delta$ are more consistent with the
feedback driven model (model-B). Finally, if the prior probability for
$\delta$ is included (Case-III, Figures~\ref{fig3}-\ref{fig4}) then we
have ruled out model-A a-priori. However in this final case it is
still instructive to investigate the consistency of model-B with
observation. The model-B values of $\delta_{\rm B}=1.66$ and
$\gamma_{\rm B}=2.5$ are consistent with the data.

These results suggest that model-B is more consistent with the data
than model-A. However it is interesting to note that the most likely
values for the parameters $\gamma$ and $\delta$ lie between those of
the two models, indicating that neither describes the astrophysics in
its totality. On the other hand, Figure~\ref{fig3} demonstrates in a
model independent way, that the data would prefer a SMBH to halo mass
ratio that is larger at higher redshifts (as suggested by the
evolution of the virial-mass---virial-velocity relation), combined
with a SMBH to halo mass ratio that is larger in higher mass
halos. These conclusions are insensitive to the addition of scatter in
the luminosity--halo mass relation, to the form of parameterisation,
and to the prior probabilities assumed.

\section{Summary}
\label{conclusion}

In this paper we have investigated the constraints that observations
of the high redshift luminosity function place on SMBH evolution. Our
approach differs from that traditionally employed. Rather than
hypothesise a physical model to govern the formation and evolution of
SMBHs, we have described a general parameterised model and determined
the regions of the resulting parameter space that are admitted by the
observations. The traditional approach allows one to demonstrate that
a particular model is viable, but does not provide a test of its
uniqueness, nor which aspects of the theory are constrained by the
data. Conversely our approach explicitly demonstrates the degeneracies
among the evolution of different physical properties, and thus reveals
which conclusions may be robustly drawn from the data.

We find that in isolation, the available data for the luminosity
function of high redshift quasars is unable to distinguish between a
range of different physical models due to a series of degeneracies
between different physical properties. The inclusion of the local SMBH
to halo mass ratio as a constraint removes some of this degeneracy and
allows us to determine the physical parameters that may most clearly
be derived from the data. We find that the mass of the high redshift
quasar host halos was $M=10^{12.5\pm0.3}M_\odot$ (90\%). The ratio of
SMBH to halo mass was larger at high redshift with a dependence
$M_{\rm bh}\propto M(1+z)^{(1.9\pm0.9)}$ [or $M_{\rm bh}\propto
Me^{(0.7\pm0.5)z}$] (90\%). In addition, the ratio of quasar
luminosity to halo mass increases with halo mass. We find $L\propto
M^{1.4\pm0.25}$ (90\%). In the instance of an accretion rate that is
insensitive to halo mass this relation is in agreement with that
observed in the local universe. In hindsight the latter result might
have been expected since the Press-Schechter mass function declines
exponentially towards high redshift, while quasar number counts are
observed to decline only as a power-law. These conclusions are not
sensitive to scatter in the relation of SMBH to halo mass, or to the
choices of parameterisation or prior probabilities, and so must be
implicit in any physical model developed to describe SMBH
evolution. However the evolution of lifetime and occupation fraction
are not currently probed by the data.

In the future, the analysis described in this paper will be enhanced
by additional constraints. For example, a larger sample of quasars
would allow the slope of the luminosity function $\beta$ and the
evolution of quasar density $B$ to be determined over a range of
redshifts. These additional constraints should break some of the
remaining degeneracies and provide us with constraints on the
evolution of quasar lifetime to go beside the current constraints
available for the host halo mass.

\section*{Acknowledgements}
The authors thank Zoltan Haiman for helpful comments on a draft of
this manuscript. The work of JSBW was supported by the Australian
Research Council.  This work was initiated when one of the authors
(TP) was visiting the School of Physics, University of Melbourne,
under the Miegunyah Distinguished Fellowship.

\label{lastpage}

\end{document}